\documentclass[secnumarabic, twocolumn, graphics,floatfix, superscriptaddress,tightenlines, aps, prb]{revtex4-2}
\usepackage[dvips]{graphicx}
\usepackage{amssymb,amsmath}
\usepackage{dcolumn}
\usepackage{bm}
\usepackage{color}
\usepackage[colorlinks=true, allcolors=blue]{hyperref}

\begin{document}

\title{
Optical cooling of nuclear spins in GaAs/(Al,Ga)As quantum wells at subkelvin temperatures: Evidence of the dynamic self-polarization of nuclear spins}

\author{M. Kotur}
\email{mladen.kotur@tu-dortmund.de}
\affiliation{Experimentelle Physik 2, Technische Universit{\"a}t Dortmund, 44227 Dortmund, Germany}

\author{D. Kudlacik}
\affiliation{Experimentelle Physik 2, Technische Universit{\"a}t Dortmund, 44227 Dortmund, Germany}

\author{N. E. Kopteva}
\affiliation{Experimentelle Physik 2, Technische Universit{\"a}t Dortmund, 44227 Dortmund, Germany}

\author{E. Kirstein}
\affiliation{Experimentelle Physik 2, Technische Universit{\"a}t Dortmund, 44227 Dortmund, Germany}

\author{D. R. Yakovlev}
\email{dmitri.yakovlev@tu-dortmund.de}
\affiliation{Experimentelle Physik 2, Technische Universit{\"a}t Dortmund, 44227 Dortmund, Germany}

\author{K. V. Kavokin}
\affiliation{Spin Optics Laboratory, St. Petersburg State University, 198504 St. Petersburg, Russia}

\author{M. Bayer}
\affiliation{Experimentelle Physik 2, Technische Universit{\"a}t Dortmund, 44227 Dortmund, Germany}
\affiliation{Research Center FEMS, Technische Universit\"at Dortmund, 44227 Dortmund, Germany}

\begin{abstract}

We investigate the dynamic polarization of nuclear spins in a nominally undoped GaAs/Al$_{0.35}$Ga$_{0.65}$As quantum well using two complementary experimental approaches: time-resolved Kerr rotation and optical orientation measurements of photoluminescence. Using the first technique, we measure a remarkably large Overhauser field of 3.1~T in a geometry close to the Faraday configuration for a 19.7~nm wide quantum well at a temperature of 1.6~K. A nuclear spin temperature of 6.4~$\mu$K is measured at an external magnetic field of 0.006~T following an adiabatic sweep from 0.6~T. Despite the quadrupole-induced nuclear spin splitting inherent to nanostructures, the nuclear spin system is found to follow the predictions of spin temperature theory. Using  the optical orientation of the photoluminescence, we investigated nuclear spin dynamics at millikelvin temperatures down to 300~mK. At a temperature of 500~mK, an Overhauser field of 160~mT is generated in an oblique but nearly Voigt magnetic field using low optical power to avoid heating. The nuclear polarization build-up time is of 150~s, consistent with earlier reports at higher temperatures, where hyperfine scattering on free photoexcited electrons governs relaxation.    At 500~mK, the onset of dynamic self-polarization of nuclear spins is observed, which becomes more pronounced as the lattice temperature is further reduced to 300~mK. The estimated nuclear spin temperature in the dynamic self-polarization regime can be as low as 200~nK.

\end{abstract}
\date{\today}

 \maketitle

\section{Introduction}
\label{sec:intro}
In recent years, the study of nuclear spin behavior in solids has gained significant renewed attention, largely driven by its potential applications in spintronics and quantum information technologies~\citep{zutic2004spintronics}. In III–V semiconductors such as GaAs, all constituent nuclei carry nonzero spin angular momentum. This characteristic gives rise to a hyperfine coupling between the spins of conduction-band electrons and the surrounding lattice nuclei, which constitutes one of the main obstacles to achieving long electron coherence times. A viable approach to mitigating this hyperfine-induced decoherence is to suppress nuclear spin fluctuations by inducing magnetic order within the nuclear spin system (NSS)~\citep{kotur2021ultra}. Since nuclear spins do not interact directly with light, the disorder of the nuclear spin system is typically reduced through nuclear spin polarization mediated by the hyperfine interaction with optically polarized electrons~\citep{meier1984optical}. Furthermore, due to the weak coupling of the NSS to the lattice, spin-polarized nuclei generally exhibit long spin relaxation times, making them attractive candidates for long-lived quantum memory~\citep{witzel2007nuclear, morton2008solid}. Thus, a thorough understanding of nuclear spin polarization and relaxation processes is essential for both fundamental studies and technological applications.
The state of an interacting nuclear spin system can be characterized by its nuclear spin temperature~\citep{meier1984optical, abragam1961principles}, which describes the population distribution of the nuclear spin states. Dynamic polarization of nuclear spins via their hyperfine interaction with optically oriented electrons allows the system to be prepared in a low-temperature state. This nuclear spin temperature can then be further reduced through adiabatic demagnetization. The initial polarization of the nuclear spins determines how low the spin temperature can ultimately be reduced~\citep{goldman1970spin}. This method provides an effective pathway for achieving deep nuclear spin cooling, potentially down to the temperature (on the order of $10^{-8}$~K) at which a phase transition to an antiferromagnetically ordered state was predicted~\citep{merkulov1982phase}.  

Unlike demagnetization cooling in metals, where the spin temperature of nuclei upon demagnetization is lowered by a reduction factor with respect to their initial spin temperature defined by precooling of the conduction electrons down to millikelvin temperatures~\citep{oja1997nuclear}, nuclear spin cooling by optical pumping does not require any precooling. The success in achieving deep nuclear spin cooling by optical pumping largely depends on the choice of a structure that enables high-degree optical polarization of nuclear spins while maintaining long spin–lattice relaxation times. However, the lattice temperature also plays an important role, reducing the electron spin relaxation via the spin-orbit interaction. In addition, the combination of low nuclear spin temperature and low lattice temperature is a prerequisite for another type of spin ordering via the formation of nuclear magnetic polarons, involving both nuclear and electron spins~\citep{merkulov_polaron, vladimirova_polaron}.

In this paper, we examine what can be achieved by combination of using a semiconductor structure with proven high efficiency of optical polarization of nuclear spins and lowering the lattice temperature down to the subkelvin temperatures. We study an undoped GaAs/(Al,Ga)As quantum well that enables efficient electron-to-nucleus spin transfer and long nuclear relaxation times~\citep{mocek2017high}, making it ideal for reaching deep nuclear spin cooling. The dynamic polarization of nuclear spins is investigated using time-resolved polarization spectroscopy based on the pump-probe Kerr rotation technique in an oblique external magnetic field, as well as by polarized photoluminescence under optical spin orientation with circular-polarized continuous-wave laser beam. The latter technique allows us to study the behavior of the electron-nuclear spin system at subkelvin temperatures, down to 300~mK.

The paper is organized as follows. In Section~\ref{sec:experiment} studied GaAs/Al$_{0.35}$Ga$_{0.65}$As structure and used experimental techniques are described. In Section~\ref{sec:TRKR} experimental results obtained by the time-resolved Kerr rotation (TRKR) technique at $T=1.6$~K are presented. The results on dynamic nuclear polarization at $T=500$~mK by means of optical orientation of photoluminescence are reported in Sec.~\ref{sec:OO}. A further decrease of the lattice temperature down to 300~mK gives strong evidence of the realization of the dynamic nuclear self-polarization presented in  Sec.~\ref{sec:auto}. Here, theoretical considerations of the dynamic nuclear self-polarization are also given.

\section{Samples and experimental setup}
\label{sec:experiment}

The high-quality sample used in this study was grown by molecular-beam epitaxy on a Te-doped GaAs substrate. It consists of 13 nominally undoped GaAs/Al$_{0.35}$Ga$_{0.65}$As quantum wells (QW) separated by 30.9~nm thick barriers. The widths of the quantum wells range from 2.8 to 39.3~nm. This sample was previously investigated in Refs.~\cite{mocek2017high,kotur2018single,kotur2021ultra, eickhoff2002coupling}. Here, we report results for a QW with width of $d=19.7$~nm. 

The hyperfine interaction between electrons and nuclei can lead to spin transfer from optically oriented electrons to nuclear spins, in a process referred to as dynamic nuclear polarization (DNP)~\cite{lampel1968nuclear, meier1984optical, dyakonov2017spin}. Polarized nuclear spins generate an effective magnetic field acting on the electrons, known as the Overhauser field $\textbf{B}_{N}$. The effect of the Overhauser field on the electron spin can be monitored via the degree of circular polarization of the photoluminescence~\citep{lampel1968nuclear} or via Kerr rotation of a linearly polarized probe beam~\citep{worsley1996transient}. In a tilted magnetic field geometry, relative to the light propagation direction, the Overhauser field of spin-polarized nuclei is collinear with the external magnetic field $\textbf{B}$. This means that the electron spin is now subject to the combined effect of the external and Overhauser fields. The sign of the Overhauser field is controlled by the helicity of the excitation light ($\sigma^+$ or $\sigma^-$). Thus, for a fixed helicity of the circularly polarized laser beam, the nuclear field can, depending on the sign of the external magnetic field, either enhance it when the two field vectors are parallel or reduce it when they are antiparallel. In this study, we used two optical techniques: time-resolved Kerr rotation and dynamical nuclear polarization induced by optically-oriented carriers and detected via polarized photoluminescence.

For the TRKR experiments the sample was mounted in a variable-temperature insert of a liquid helium bath cryostat, allowing operation from 1.6~K up to room temperature. At $T=1.6$~K the sample was in direct contact with superfluid helium. The cryostat was equipped with a superconducting vector magnet capable of applying magnetic fields up to 3~T in any direction. In our experiments, the magnetic field $\mathbf{B}$ was applied either in the Faraday geometry ($\theta = 0^\circ$, $B_\parallel \parallel \mathbf{k}$), Voigt geometry ($\theta = 90^\circ$, $B_\perp \perp \mathbf{k}$), or at an intermediate tilt angle $\theta$ between the two ($\theta = \angle(\mathbf{B}, \mathbf{k})$). Here, $\mathbf{k}$ is the wave vector of the pump laser, aligned with the sample growth axis ($z$-axis). The TRKR technique~\citep{dyakonov2017spin} was used to study nuclear spin polarization and evaluate the Overhauser field via its influence on electron spins. Excitation with a circularly polarized pump pulse generates electron spins oriented along the light propagation axis. Applying the external magnetic field in an oblique geometry causes the initial spin polarization to precess about $\textbf{B}$ at the Larmor precession frequency~\citep{kikkawa2000all}  
\begin{equation}
\nu_L=\frac{|g_e| \mu_B B}{h} \,.
\label{eq:larmor_frequency}
\end{equation}
Here $g_e$ is the electron $g$-factor, $\mu_B$ is the Bohr magneton, and $h$ is the Planck constant. A time-delayed, linearly polarized probe pulse monitors the evolution of the spin projection along the growth axis, $S_z$, via the Kerr rotation $\theta_K$, since $\theta_K \propto S_z$. For the analysis of the TRKR signal, the data were fitted using the following equation, characteristic of Larmor precession in an external magnetic field~\citep{dyakonov2017spin}
\begin{equation}
\theta_{\rm K}(t)=A \exp{(-t/T_2^*)} \cos(2\pi \nu_{\rm L} t+\phi).
\label{eq:TRKR_fit}
\end{equation}
Here, $A$ is the amplitude of the TRKR signal, $T_2^*$ is the spin dephasing time and $\phi$ is the initial phase of the oscillation.

\begin{figure*}
\center{\includegraphics[width=0.54\textwidth]{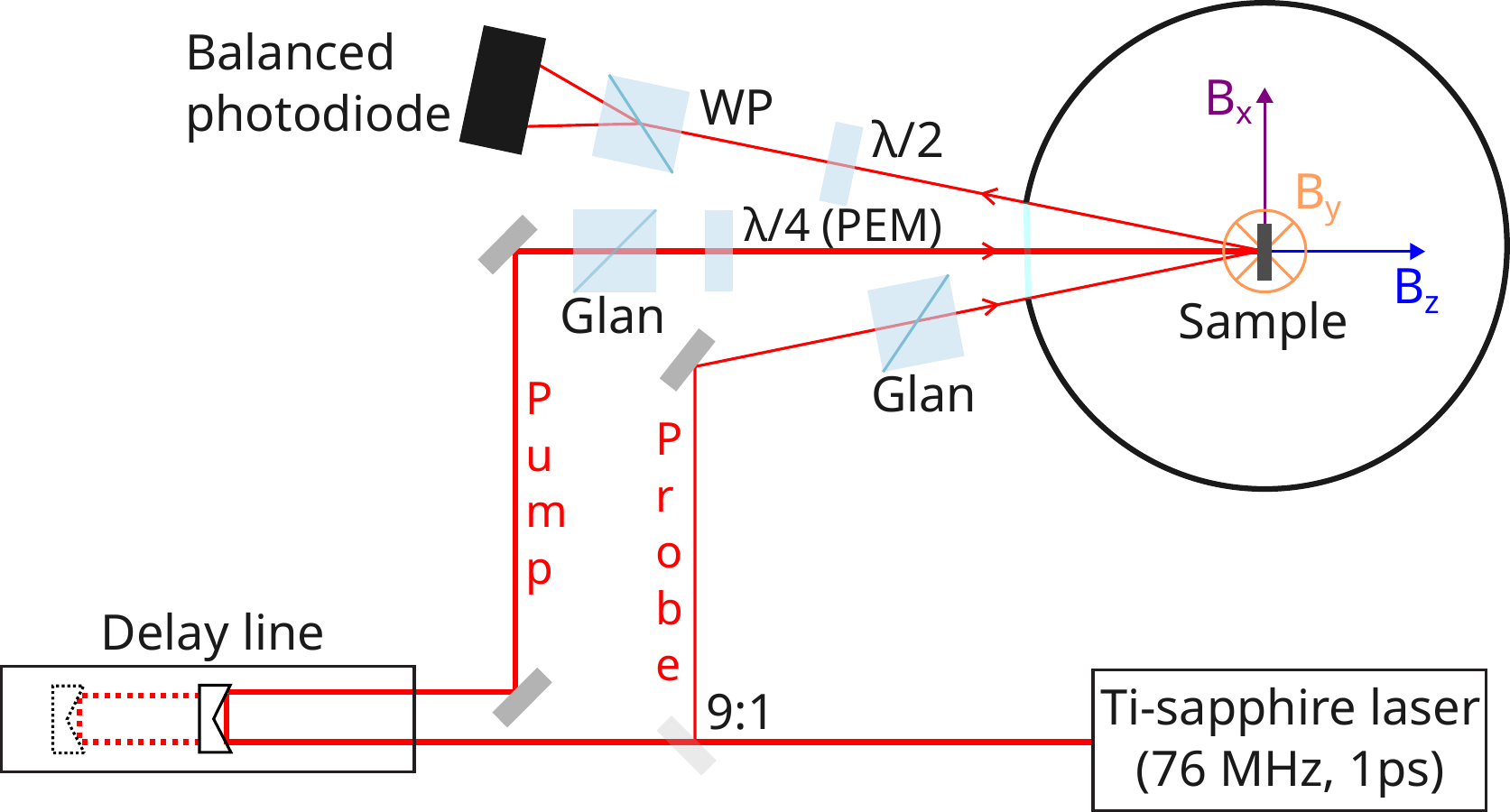}}
\caption{Time-resolved Kerr rotation setup. Magnetic field components $B_x$, $B_y$ and $B_z$ indicate field orientations corresponding to Voigt and Faraday geometries. For a circularly polarized pump with constant $\sigma^+$ helicity, a $\lambda/4$ wave plate was used, which was replaced by a photoelastic modulator (PEM) for a modulated $\sigma^+/\sigma^-$ pump.}
\label{fig:setup_TRKR}
\end{figure*}

The TRKR experimental setup is schematically shown in Fig.~\ref{fig:setup_TRKR}. A mode-locked Ti:Sapphire laser with a repetition rate of 76 MHz provided the pulsed excitation, with a pulse duration of 1.5 ps and a spectral half-width at half-maximum of about 1 meV. The output beam was split into a circularly-polarized pump used to generate the spin-polarized electrons and a time delayed linearly-polarized probe. The pump polarization was either kept constant $\sigma^+$ using a $\lambda/4$ wave plate or modulated $\sigma^+ / \sigma^-$ by a photoelastic modulator (PEM). The reflected probe beam was detected using a balanced photodiode after polarization components of an achromatic half waveplate and a Wollaston prism (WP) for Kerr rotation analysis.

\begin{figure*}
\center{\includegraphics[width=0.7\textwidth]{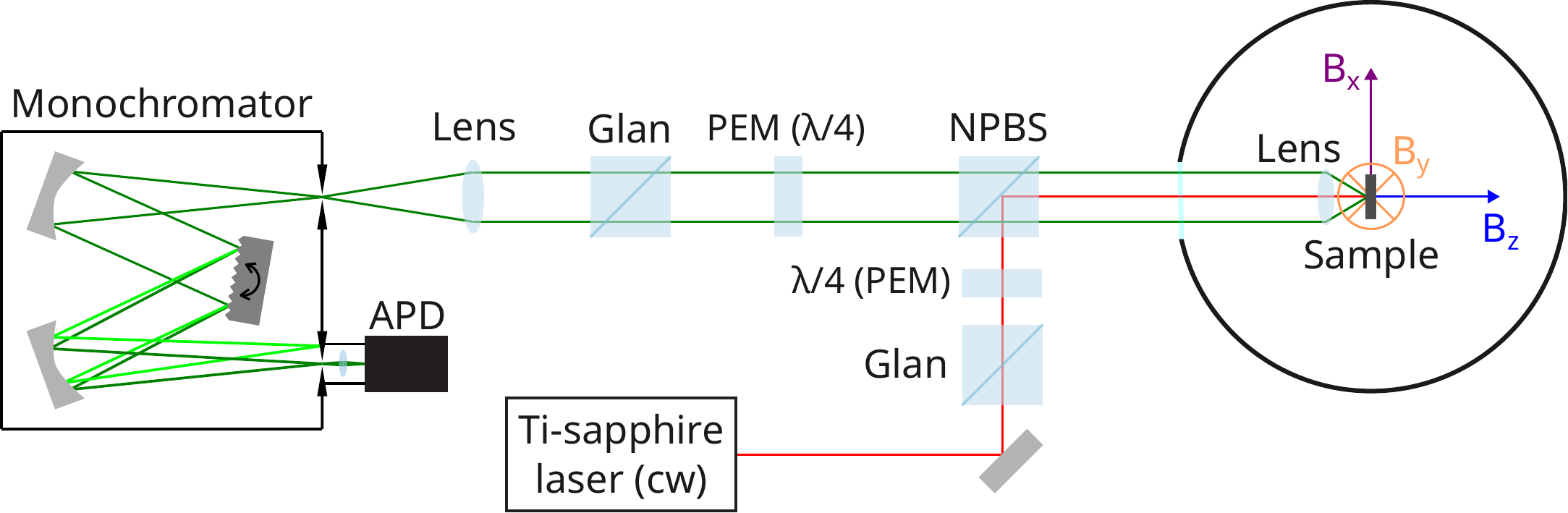}}
\caption{Setup for DNP and optical nuclear cooling. NPBS is non-polarizing beam splitter. The magnetic field components $B_x$, $B_y$ and $B_z$ indicate field orientations corresponding to Voigt and Faraday geometries.}
\label{fig:setup_mK}
\end{figure*}

For the DNP and nuclear cooling experiments at temperatures below 1.6~K, the sample was mounted on a copper puck thermally anchored to the mixing chamber plate of a cryogen-free dilution refrigerator (Oxford Instruments Proteox), with the base temperature reached without sample illumination of $T=10\pm2$~mK. The cryostat is equipped with optical windows, allowing direct optical excitation and detection. The temperature was monitored using two calibrated sensors, one positioned on the mixing plate ($T_m$) and another one placed on the sample holder in close proximity to the sample ($T_s$). The cryostat is equipped with a superconducting vector magnet, generating fields up to $B_{z}=5$~T in the Faraday geometry (field is parallel to the light wave vector) and up to $B_{x}=B_{y}=1$~T in the Voigt geometry. A lens was placed inside the puck just next the sample and aligned with the optical axis of the window, see Fig.~\ref{fig:setup_mK}. It was used to focus the excitation laser beam onto the sample  and to collect the photoluminescence (PL). To photogenerate spin-polarized electrons, the sample was excited with circularly polarized light ($\sigma^+$) from a Ti:Sapphire laser tuned to $E_{exc}=1.535$~eV. PL was detected using an avalanche photodiode connected to a photon counter, following spectral selection by a monochromator. The circular polarization components were analyzed using a polarization analyzer consisting of a photoelastic modulator (PEM) and a Glan-Taylor prism. The PEM was synchronized with the photon counter to enable polarization-resolved detection. The intensities of the $\sigma^+$ ($I^+$) and $\sigma^-$ ($I^-$) circularly polarized emission were recorded, allowing us to measure the optical orientation degree: $P_{\rm oo}=(I^+-I^-)/(I^++I^-)$~\cite{meier1984optical,dyakonov2017spin}. 

\section{Experimental Results}
\label{sec:results}

\begin{figure}
\center{\includegraphics{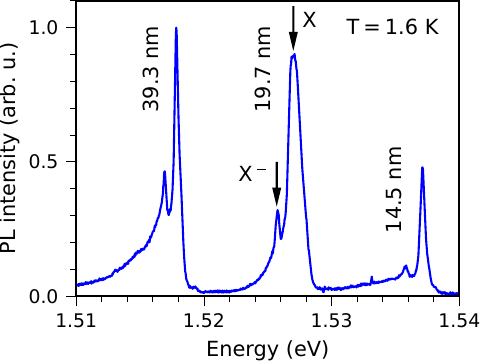}}
\caption{Photoluminescence spectrum of the three widest out of thirteen GaAs/Al$_{0.35}$Ga$_{0.65}$As quantum wells measured at $T = 1.6$~K. The widths of the quantum wells are shown near their respective spectral lines, while arrows mark the position of the exciton (X) and trion (X$^-$) lines for the 19.7-nm-wide quantum well. Excitation energy $E_{exc} = 1.823$~eV and power $P = 1$~mW.}
\label{fig:spectrum_all}
\end{figure}

The photoluminescence spectrum of the GaAs/Al$_{0.35}$Ga$_{0.65}$As sample containing thirteen quantum wells with widths ranging from 2.8 to 39.3~nm, measured at $T = 1.6$~K, is shown in Fig.~\ref{fig:spectrum_all} for the three widest quantum wells. All measurements presented in this work were conducted on the 19.7 nm-wide QW. The PL spectrum of this QW consists of two emission lines at 1.527 and 1.526~eV, the first being attributed to exciton ($X$) and the second to trion ($X^-$) \citep{mocek2017high}. The presence of the trion line, despite the sample being nominally undoped, indicates the presence of a small concentration of resident electrons ($n_e \leq 10^{10}~\text{cm}^{-2}$).

\subsection{Nuclear cooling detected via TRKR  at 1.6 K}
\label{sec:TRKR}

Time-resolved Kerr rotation was used to study DNP and measure the resulting Overhauser field via its effect on the coherent dynamics of electron spins. Measurements were performed at $T=1.6$~K, with the Ti:Sapphire laser set to $E_{pump}=1.527$~eV. The pump was circularly polarized with $P_{pump}=10$~mW (unless stated otherwise), and the probe was linearly polarized  with $P_{probe}=0.5$~mW. When the sample is excited with helicity modulated light $\sigma^+ / \sigma^-$, the nuclei remain unpolarized. In this case, the electron spins feel only the external magnetic field. By scanning the pump-probe delay and measuring how the electron spin projection along the $z$ axis, $S_{z}$, oscillates, the electron $g$-factor can be determined from Eq.~\eqref{eq:larmor_frequency}. In contrast, for constant $\sigma^+$ excitation, DNP becomes possible, and the electrons are subjected to the combined action of the external and Overhauser magnetic fields. 

\begin{figure}
\center{\includegraphics{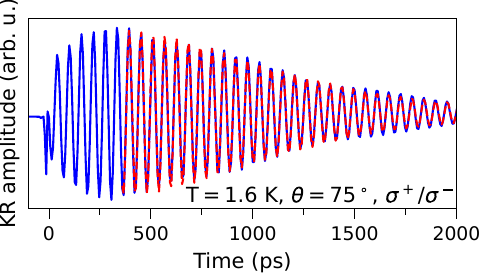}}
\caption{Time-resolved Kerr rotation signals measured at $T = 1.6$~K in a magnetic field of $B = 3$~T tilted by $\theta = 75^\circ$. The pump was modulated between $\sigma^+$ and $\sigma^-$ polarization at 50 kHz, with energy $E_{pump} = 1.527$~eV and power $P_{pump} = 10$~mW. The red line is a fit to Eq.~\eqref{eq:TRKR_fit}.}
\label{fig:75deg}
\end{figure}

\begin{figure}
\center{\includegraphics{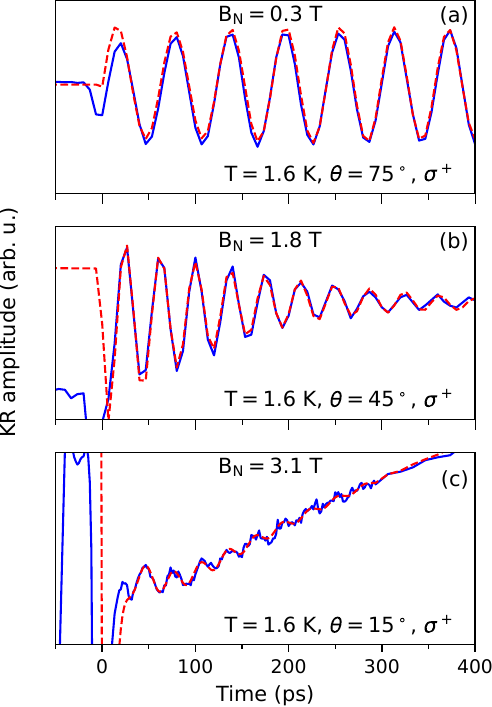}}
\caption{Comparison of TRKR signals measured at $T = 1.6$~K in tilted magnetic field of $B = 3$~T at: (a) $\theta = 75^\circ$, (b) $\theta = 45^\circ$ and (c) $\theta = 15^\circ$. The pump has constant $\sigma^+$ circular polarization, $E_{pump} = 1.527$~eV, and $P_{pump} = 10$~mW. The red lines are fits using Eq.~\eqref{eq:TRKR_fit} with parameters given in Table~\ref{tab:nuclear_fields}.}
\label{fig:15degAnd45deg}
\end{figure}

Figure~\ref{fig:75deg} shows the Kerr rotation amplitude measured for the modulated $\sigma^+ / \sigma^-$ pump in an external magnetic field of $B = 3$~T tilted by $\theta = 75^\circ$ from the $z$-axis. In this case, the measured Larmor precession frequency is $\nu_L = 17.0$~GHz, which corresponds to $g_e = -0.40$. This value is in good agreement with the previously reported value for a 20-nm-wide QW and is close to that of bulk GaAs $g_e = -0.44$~\citep{snelling1991magnetic}. For a constant $\sigma^+$ excitation, the Larmor precession frequency measured at $B=3$~T increases to $\nu_L = 18.1$~GHz, corresponding to a total field of $B+B_N=3.3$~T and revealing the nuclear Overhauser field of $B_N=0.3$~T, see Fig.~\ref{fig:15degAnd45deg}(a).

In order to increase the DNP and the Overhauser field, we performed measurements at smaller angles of $\theta = 45^\circ$ and $15^\circ$. The resulting TRKR dynamics for constant $\sigma^+$ pump polarization are shown in Figs.~\ref{fig:15degAnd45deg}(b,c). The external magnetic field was maintained at $B = 3$~T. To minimize the effect of the linearly polarized probe beam, which can perturb the nuclear spin system via inelastic scattering~\citep{aleksandrov2011spin}, the probe was blocked for 30 minutes prior to measuring the TRKR signal at an angle of $\theta = 15^\circ$. The obtained Larmor precession frequencies $\nu_L$ and the corresponding Overhauser fields  are given in Table~\ref{tab:nuclear_fields}.
\begin{table}
\centering
\caption{Evaluated Larmor precession frequencies and corresponding Overhauser fields for various tilt angles of the external magnetic field $B=3$~T. $P_N$ is the nuclear polarization degree evaluated from measured $B_N$ for $g_e=-0.40$ and maximum Overhauser field of 4.3~T corresponding to $P_N=100\%$ (see text).}
\begin{tabular}{|c|c|c|c|} \hline
Angle $\theta$ ($^\circ$) &  $\nu_L$ (GHz) & $B_N$ (T) & $P_N (\%)$ \\ \hline\hline
75 & 18.1 & 0.3 &   7 \\ \hline
45 & 26.7 & 1.8 &  42 \\ \hline
15 & 34.2 & 3.1 &  72 \\ \hline
\end{tabular}
\label{tab:nuclear_fields}
\end{table}
One can see, that for $\theta=15^\circ$ the Overhauser field is $B_N=3.1$~T. Note, that similar or even larger values, up to 4~T, for the Overhauser field have been reported in GaAs quantum dots, where electron confinement leads to a much stronger interaction with nuclei than in the bulk material~\cite{baugh2007large, tartakovskii2007nuclear}. In the studied 19.7-nm-wide GaAs quantum well we reach the Overhauser field of 3.1~T for $\theta =15^\circ$.  The full nuclear polarization of 100\% corresponds to the state in which all Ga and As nuclear spins are fully aligned, resulting in the electron spin splitting of $\Delta_{eN}\approx100$~$\mu$eV~\citep{millington2024approaching}. With the electron $g$-factor $g_e=-0.4$, this corresponds to the maximum Overhauser field
\begin{equation}
B_{N}^{max}=\frac{\Delta_{eN}}{\mu_{B}|g_e|} \,
\label{eq:B_Nmax}
\end{equation}
equal to 4.3~T. Therefore, the Overhauser field $B_N=3.1$~T corresponds to $72$\% of the full nuclear polarization reached in our structure.  Considering the weak confinement in such a wide quantum well, this result represents a surprisingly efficient nuclear spin polarization compared to the typically stronger effects observed in quantum dots. Note that the nuclear spin ordering effects in semiconductors are predicted to occur once the nuclear polarization reaches 70\%. At this polarization level, the NSS system may undergo a phase transition into a ferromagnetic or antiferromagnetic state~\citep{merkulov1982phase} upon adiabatic demagnetization to zero external field. 

To relate the measured Overhauser fields to the nuclear spin temperature, we employ a spin cooling procedure based on the principle of adiabatic demagnetization~\citep{vladimirova2018spin}. The nuclear cooling procedure begins with pumping the nuclear spins in a longitudinal magnetic field of $B_{\parallel}=0.6$~T for 30~min. During this time, the pump laser power is set to $P_{pump}=10$~mW, and the probe beam is blocked. After 30 min, the pump beam is also blocked, and the magnetic field is gradually rotated from the Faraday to the Voigt configuration, keeping its magnitude constant at 0.6~T. Then the Voigt field is slowly reduced from its initial value $B_\perp(0)$ (in our case $B_\perp(0)=0.6$~T) to $B_\perp$, such that $dB_{\perp}/dt<B_L/T_2$. By this adiabatic demagnetization the nuclear spin temperature $\Theta_N$ follows the universal expression dictated by entropy conservation during adiabatic demagnetization
\begin{equation}
\Theta_N=\Theta_N(0)\sqrt{\frac{B_{\perp}^2+B_L^2}{B_{\perp}(0)^2+B_L^2}}.
\label{eq:spintemperature}
\end{equation}
Here $\Theta_N(0)$ is the initial nuclear spin temperature before the adiabatic demagnetization prepared in the external magnetic field $B_{\perp}(0)$ and $T_2$ is the transverse spin relaxation time, which is determined by spin–spin interactions between nuclei. In GaAs it is on the order of 100~$\mu$s~\citep{meier1984optical}. $B_L$ is the local magnetic field, which is about 0.15~mT in bulk GaAs~\citep{paget1977low}. This field originates from the dipole–dipole interaction between neighboring nuclei, which creates a local magnetic field at each nuclear site. In nanostructures, this field can be significantly enhanced due to strain effects inherent to these systems~\citep{vladimirova2017nuclear}. Accordingly, the nuclear spin polarization $P_N$ can be expressed as
\begin{equation}
P_N=\frac{B_{\perp}}{3k_B\Theta_N}\hbar \langle \gamma_N(I+1) \rangle.
\label{eq:spinpolarization}
\end{equation}
Here, $I$ is the nuclear spin, $k_B$ is the Boltzmann constant and $\gamma_N$ is the nuclear gyromagnetic ratio. The angular brackets denote averaging over all nuclear species ($^{69}$Ga, $^{71}$Ga and $^{75}$As) weighted by their natural abundances~\cite{vladimirova2018spin}. 

\begin{figure}
\center{\includegraphics{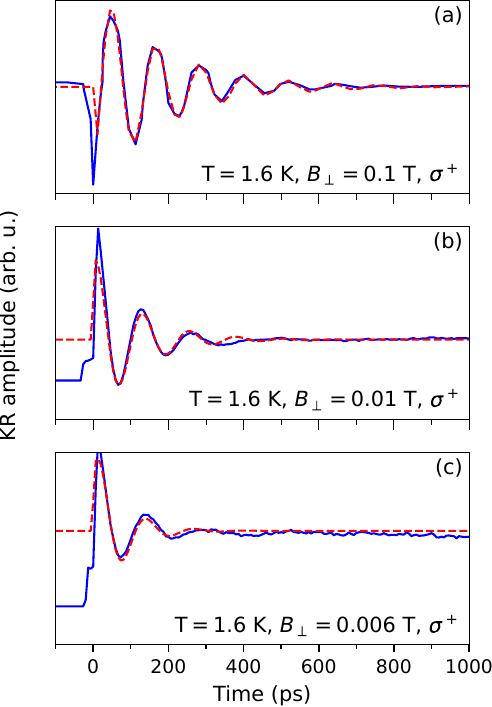}}
\caption{Time-resolved Kerr rotation dynamics for three Voigt magnetic fields ($P_{pump} = P_{probe} = 0.5$~mW). Prior to measurement, the nuclear spin system (NSS) was pumped for 30~min in $B_{\parallel}=0.6$~T  ($P_{pump} = 10$~mW, probe off). After pumping, the field was rotated to the Voigt geometry and reduced to the values given in panels. Experimental data are shown by solid blue lines. Dashed red lines are fits using Eq.~\eqref{eq:TRKR_fit} with parameters given in Table~\ref{tab:cooling_parameters}.}
\label{fig:nst}
\end{figure}

When the external field is decreased during the adiabatic demagnetization to the desired target value, the pump, whose power is reduced to $P_{pump}=0.5$~mW, and the probe beams ($P_{probe}=0.5$~mW) are unblocked and the TRKR signal oscillations are recorded. The data obtained for $B_{\perp}=0.1$, 0.01, and 0.006~T are shown in Fig~\ref{fig:nst}. From the extracted Larmor frequencies $\nu_L$ of 8.5, 8.0, and 7.8~GHz, the Overhauser field $B_N$ is found to be about 1.4~T, independent of the $B_{\perp}$ values. Such behavior is expected from Eqs.~\eqref{eq:spinpolarization} and \eqref{eq:spintemperature} for external magnetic fields which are much stronger than the local magnetic field, $B_{\perp} \gg B_L$. The evaluated nuclear spin temperatures from Eq.~\eqref{eq:spinpolarization} for the three external magnetic fields $B_{\perp}=0.1$, 0.01 and 0.006~T are $\Theta_N=106.5$, 10.7 and 6.4~$\mu$K, respectively. It follows from Eq.~\eqref{eq:spintemperature} that for $B_{\perp}=0$~T the nuclear spin temperature would be $\Theta_N=0.8$~$\mu$K for $B_L=0.8$~mT~\citep{mocek2017high}. The larger value of $B_L$ in the quantum well, compared to bulk GaAs where it is governed by dipole-dipole interactions between nuclei, arises from the additional strain inherent to micro- and nanostructures.

Unfortunately, due to difficulties with compensating the parasitic residual fields in our superconducting magnet, it was not possible to decrease the external field below 6~mT. For comparison, on the same sample, we previously measured a nuclear spin temperature of 0.54~$\mu$K using the adiabatic demagnetization in the rotating frame (ADRF) method~\citep{kotur2021ultra}. Although lower nuclear spin temperatures have been reported for CdTe quantum wells, $\Theta_{ N}=0.3$~$\mu$K~\citep{gribakin2024nuclear}, $\Theta_{N}=0.54$~$\mu$K remains, to the best of our knowledge, the lowest measured nuclear spin temperature for GaAs.

\begin{table}
\centering
\caption{Parameters for Fig.~\ref{fig:nst}.}
\begin{tabular}{|c|c|c|c|c|} \hline
$B_\perp$ (T) & $\nu_{L}$ (GHz) &  $B_N$ (T) & $\Theta_N$ ($\mu$K) & $T_2^*$ (ps) \\ \hline\hline
0.1 & 8.5 & 1.4 & 106.5 & 190 \\ \hline
0.01 & 8.0 & 1.4 & 10.7 & 110 \\ \hline
0.006 & 7.9 & 1.4 & 6.4 & 70\\ \hline
\end{tabular}
\label{tab:cooling_parameters}
\end{table}

\subsection{Nuclear spin cooling detected by optical orientation of electrons at 500~mK}
\label{sec:OO}

\begin{figure}
\center{\includegraphics{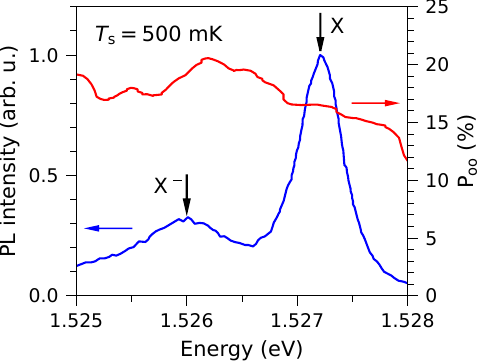}}
\caption{Photoluminescence spectrum (blue line) of the 19.7-nm-wide GaAs/Al$_{0.35}$Ga$_{0.65}$As QW measured at $T_s = 500$~mK. Excitation energy $E_{exc} = 1.535$~eV and power $P=0.1$~mW. X and X$^-$ denote exciton and trion lines. The red line shows spectral dependence of the optical orientation degree.}
\label{fig:spectrum}
\end{figure}

Figure~\ref{fig:spectrum} presents the PL spectrum of the 19.7-nm-wide QW measured at $T_s=500$~mK, as indicated by the temperature sensor at the sample. Although the base temperature of our cryostat system is about 10~mK, optical excitation causes heating of the sample, which is monitored with a sensor placed in close proximity. To this end, a laser power of 0.1~mW is used as a compromise between limiting sample heating, keeping the temperature in the millikelvin range, and maintaining a sufficient signal-to-noise ratio. The maximum intensity of the PL at 1.527~eV, attributed to exciton (X) emission, was chosen as the PL detection energy for all subsequent measurements. Figure~\ref{fig:spectrum} also shows the spectral dependence of the optical orientation $P_{\rm oo}$, measured for modulated circularly polarized light $\sigma^+ / \sigma^-$ at 50~kHz. At the exciton emission energy, 1.527~eV, $P_{\rm oo}$ reaches a value of about 17\%.

In order to examine the DNP and optical cooling of the nuclear spin system at temperatures below 1.6~K we use the optical orientation technique detecting the polarization of photoluminescence. The Hanle effect is used for that, which  describes the depolarization of luminescence by a transverse external magnetic field~\citep{hanle1924magnetische}, or, in our case, by the transverse component of a field applied at an angle. If the sample is excited  with helicity modulated light ($\sigma^+/\sigma^-$ at 50 kHz), due to the rapid changes in electron spin, the nuclei remain unpolarized~\citep{meier1984optical}. In this case the electron spin is influenced only by the external magnetic field and the Hanle curve is symmetric with respect to magnetic field inversion. 

\begin{figure}
\center{\includegraphics{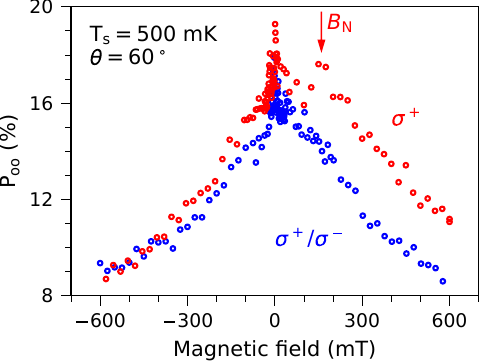}}
\caption{Hanle curve in a magnetic field tilted by $\theta = 60^\circ$ measured with modulated $\sigma^+/\sigma^-$ (blue circles) and constant $\sigma^+$ (red circles) excitation helicity at $T_s = 500$~mK. $E_{exc} = 1.535$~eV and $P=0.1$~mW. The arrow indicates the strength of the Overhauser field of spin polarized nuclei.}
\label{fig:hanle_tilted}
\end{figure}

The Hanle curve measured under helicity modulated excitation at a temperature of $T_s=500$~mK ($T_m=45$~mK) is shown in Fig.~\ref{fig:hanle_tilted}, indicated by blue circles. The external magnetic field $\textbf{B}$ is applied at an angle $\theta=60^\circ$. Due to the absence of the Overhauser field, the Hanle curve exhibits symmetry about zero field. The Hanle curve changes considerably for $\sigma^+$ excitation (red circles in  Fig.~\ref{fig:hanle_tilted}), with a shoulder appearing at positive magnetic fields and a maximum at 0.16~T. At this magnetic field, the Overhauser field exactly compensates the external one, therefore, one can evaluate $B_N=-0.16$~T. The steady-state nuclear magnetic field $B_N^{st}$, i.e., the effective magnetic field produced by nuclei once their polarization has reached equilibrium under optical pumping, is given by~\citep{d1974optical, mocek2017high}
\begin{equation}
B_N^{st}=B_N^{max}  \frac{I+1}{S(S+1)}  f  S_0  \text{cos}\theta.
\label{eq:nuclear_steady_state}
\end{equation}
Here, $B_N^{max}=4.3$~T is the maximum Overhauser field defined by Eq.~\eqref{eq:B_Nmax}, and $S_0$ is the mean electron spin. Since we detect the PL emission due to recombination of electrons with heavy holes having the spin projection on the QW axis equal to $\pm1/2$, $S_0$ can be evaluated as $S_0=P_{\rm oo}/2$. The parameter $f$ in Eq.~\eqref{eq:nuclear_steady_state} is the leakage factor, defined as
\begin{equation}
f=\frac{T_L}{T_{1e}+T_L},
\label{eq:leakage_factor}
\end{equation}
where $T_{1e}$ and $T_L$ denote the nuclear spin relaxation times via electrons and via all other channels, respectively. The leakage factor effectively quantifies the efficiency of dynamic nuclear polarization driven by optically oriented electrons. Since in our case $|B_{N}^{st}|=0.16$~T (corresponding to $P_N=0.037$) and $\theta=60^\circ$, it follows from Eq.~\eqref{eq:nuclear_steady_state} that the leakage factor is $f \approx 0.26$. It is worth nothing that the nuclear spin polarization $P_N=0.037$ iss achieved at a very low excitation power of $P=0.1$~mW, which we kept low to maintain the temperature in the millikelvin range. Increasing the excitation density would increase $P_{N}$ and, consequently, the Overhauser field $B_{N}$, but it would also cause lattice heating, which has the opposite effect on $P_{N}$. For comparison, in our previous work~\citep{kotur2018single} we were able to reach Overhauser field values close to 0.6~T (corresponding to $P_N=0.14$) at $T=1.8$~K, but with a much higher excitation power of $P=10$~mW.

To gain deeper insight into the nuclear spin behavior, we proceeded with an investigation of the nuclear spin dynamics under illumination. To this end, we followed a protocol analogous to those described in Refs.~\citep{kalevich1982onset, kotur2014nuclear}. In the first stage, the circularly polarized ($\sigma^+$) excitation light is blocked for 500~s using a mechanical shutter, and no external magnetic field was applied. This ensured that, at the beginning of the second stage, there is no nuclear spin polarization, as the spins relax during the dark period with a characteristic spin-lattice relaxation time $T_1$, where $T_1^{-1}=T_{1e}^{-1}+T_L^{-1}$. For the same 19.7-nm-wide QW, the relaxation time $T_1$ during optical pumping depends on temperature and is about 400~s at $T=10$~K~\citep{mocek2017high}, decreasing to just above 200~s at $T=1.8$~K~\citep{kotur2018single}. In contrast, at $T=10$~K, in darkness and at zero magnetic field, $T_1$ is only a few seconds~\citep{mocek2017high}. Therefore, the chosen 500 s dark period is sufficiently long to ensure that no nuclear polarization remains at the start of the second stage. 

\begin{figure}
\center{\includegraphics{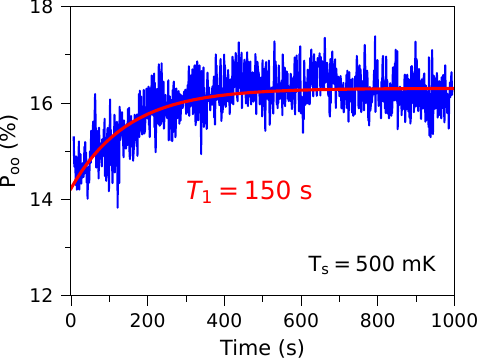}}
\caption{Build-up of nuclear spin polarization measured via the change in the optical orientation degree after the sample was exposed to illumination, following a 500~s dark period, in a magnetic field of $B = 0.16$~T tilted by $\theta = 60^\circ$. Sample temperature is $T_s = 500$~mK, with excitation energy $E_{exc} = 1.535$~eV and excitation power $P = 0.1$~mW. Red line shows an exponential fit, from which the nuclear spin polarization build-up time $T_1=150$~s is evaluated.}
\label{fig:relaxation}
\end{figure}

Just before the end of the first stage, a tilted  magnetic field of $0.16$~T is applied at an angle of $\theta=60^\circ$. The laser is then unblocked, and the optical orientation degree is recorded as a function of time. The measured $P_{\rm oo}$ dynamics is shown in Fig.~\ref{fig:relaxation}. It can be clearly seen that the degree of optical orientation increases with time and reaches saturation. The reason for this is that during excitation, nuclear spins become dynamically polarized, resulting in the emergence of the Overhauser field. This field compensates the external magnetic field, reducing the effective field on the electrons and thereby increasing the degree of optical orientation. The time dependence of the Overhauser field is given by~\citep{mocek2017high}
\begin{equation}
B_{N}(t)=B_{N}^{st} [1-\exp{(-t/T_1)}].
\end{equation}
Thus, by fitting the $P_{\rm oo}(t)$ dynamics with an exponential function $\sim \exp{(-t/T_1)}$, we obtain the $T_1$ time under illumination. The experimental value extracted from Fig.~\ref{fig:relaxation} is $T_1=150$~s. This time is compared with previously measured values on the same sample at temperatures ranging from 1.8 to 20~K, as summarized in Table~\ref{tab:relaxation_times}. For temperatures above 6~K, the relaxation time $T_1$ remains temperature-independent, despite some scattering of the experimental data. As reported in Ref.~\citep{mocek2017high}, the dominant relaxation mechanism of nuclear spins is hyperfine scattering on free photoexcited two-dimensional electrons. This process is described theoretically in Ref.~\citep{kalevich1990optical}, where the spin relaxation rate for nondegenerate free electrons in a QW of width $d$ was derived as
\begin{equation}
\frac{1}{T_{1,e}} \propto \frac{A^2 \Omega^2 n_e m^*}{\hbar^3 d^2}.
\label{eq:relaxation}
\end{equation}
Here, $A$ denotes the hyperfine constant, $\Omega$ represents the unit cell volume, $n_e$ is the two-dimensional electron concentration,  and $m^*$ is the effective electron mass.

Equation~\eqref{eq:relaxation} indicates that the spin relaxation time $T_{1,e}$ should remain invariant with respect to temperature. However, at temperatures below 6~K, it is observed to decrease as the temperature is lowered. A plausible explanation for the $T_1$ temperature dependence observed below 6~K is the onset of electron localization~\citep{kotur2018single}. In this regime, localized electrons, owing to their longer correlation time, transfer spin to the nuclei more efficiently than free electrons, thereby accelerating nuclear spin relaxation~\citep{meier1984optical}.

\begin{table}
\centering
\caption{Comparison of the nuclear spin-lattice relaxation time measured at $T_s = 500$~mK with previously reported values at Kelvin-range temperatures for the 19.7-nm-wide GaAs/Al$_{0.35}$Ga$_{0.65}$As QW.}
\begin{tabular}{|c|c|} \hline
Temperature (K) & Relaxation time $T_1$ (s) \\ \hline\hline
0.5 & 150 \\ \hline
1.8 & 210~\citep{kotur2018single} \\ \hline
6 & 300~\citep{kotur2018single} \\ \hline
10 & 450~\citep{kotur2018single} \\ \hline
10 & $\sim400$~\citep{mocek2017high} \\ \hline
20 & $\sim300$~\citep{mocek2017high} \\ \hline
\end{tabular}
\label{tab:relaxation_times}
\end{table}

\subsection{Dynamic self-polarization of nuclear spins at 300~mK}
\label{sec:auto}

\begin{figure*}
\center{\includegraphics{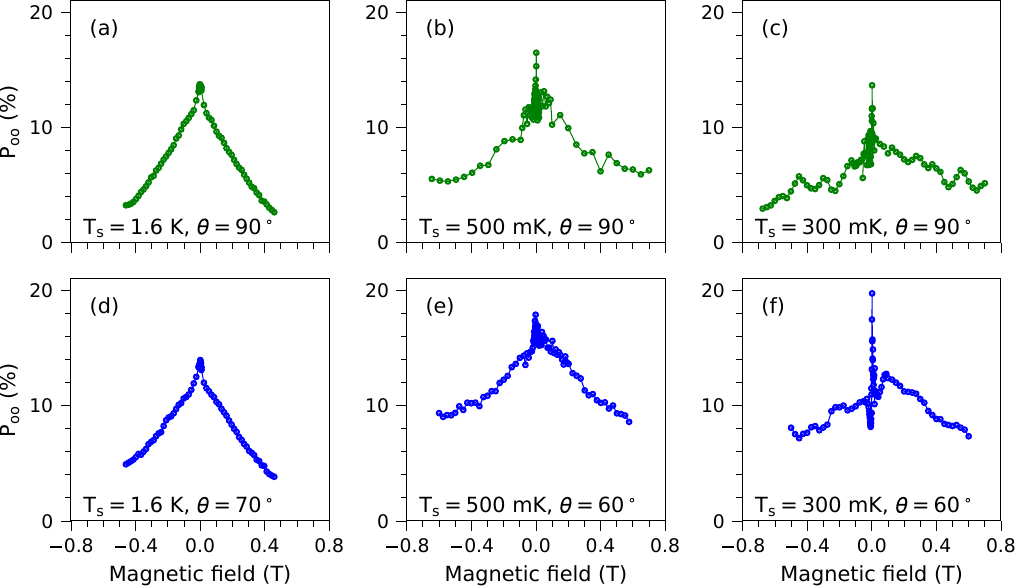}}
\caption{Hanle curves measured in a Voigt magnetic field ($\theta = 90^\circ$) under modulated $\sigma^+/\sigma^-$ excitation with $E_\mathrm{exc} = 1.535$ eV and (a) $T_s = 1.6$ K and $P = 0.5$ mW, (b) $T_s = 500$ mK and $P = 0.1$ mW, and (c) $T_s = 300$ mK and $P = 0.01$ mW. Hanle curves measured in a tilted magnetic field geometry under modulated $\sigma^+/\sigma^-$ excitation for (d) $T_s = 1.6$ K, $P = 0.5$ mW, and $\theta = 70^\circ$, (e) $T_s = 500$ mK, $P = 0.1$ mW, and $\theta = 60^\circ$, and (f) $T_s = 300$ mK, $P = 0.01$ mW, and $\theta = 60^\circ$.}
\label{fig:Hanle_60_90_deg_Temperature}
\end{figure*}

In order to examine the properties of the electron-nuclear spin system at the minimal lattice temperatures, where the regime of the dynamic self-polarization of nuclear spins (DSPNS) could be reached~\citep{dyakonov1972dynamic}, we further reduced the excitation power to 0.01~mW. Under this condition the sample temperature of $T_{s} = 300$~mK is reached. We use the optical orientation technique with detection of the circular polarization of PL, as described in Sec.~\ref{sec:OO}. The Hanle curves measured under helicity-modulated circular excitation ($\sigma^+/\sigma^-$) are shown in Fig.~\ref{fig:Hanle_60_90_deg_Temperature} for perpendicular ($\theta=90^\circ$) and tilted ($\theta=60^\circ$) magnetic fields with respect to the light wave vector $\mathbf{k}$, at sample temperatures of 500~mK and 300~mK. 
For comparison, measurements performed at 1.6~K, when the sample is in direct contact with liquid helium, are also shown in Fig.~\ref{fig:Hanle_60_90_deg_Temperature}(a,d) for $\theta=90^\circ$ and $70^\circ$.

The main qualitative feature of the Hanle curves, that appears when cooling the structure below 1~K, is a narrow peak at zero magnetic field. One can see by comparing the data at 300~mK and 500~mK that the $P_{\rm oo}$ value at $B=0$ does not change significantly, but the wings at finite magnetic fields decrease strongly with decreasing temperature, so that the narrow peak becomes much more pronounced.

The appearance of a zero-field peak is usually considered as a fingerprint of the dynamic polarization of nuclear spins by spin-polarized electrons. It reflects building up the Overhauser field, that becomes possible when the external field overcomes local fields of nuclear spin-spin interactions~\citep{dyakonov2017spin, meier1984optical}. However, in our case the electron spin polarization is alternated at a high frequency, with the period much shorter than both spin-spin and spin-lattice relaxation times of nuclei (20~$\mu$s  vs approximately 0.1~ms and 100~s, respectively). Under these conditions, spin polarization cannot accumulate in the nuclear spin system~\citep{meier1984optical}, so that, regarding their interaction with nuclear spins, the electrons are effectively unpolarized.

Dynamic polarization of nuclei by unpolarized photoexcited electrons via the Overhauser effect~\citep{meier1984optical} is indeed possible at ultra-low temperatures, but it requires application of an external magnetic field, which is close to zero in the vicinity of the zero-field peak. Dyakonov and Perel suggested that the Overhauser effect may occur in the effective magnetic field of polarized nuclear spins (the Overhauser field) and proposed a theoretical prediction of the dynamic self-polarization of nuclear spins at temperatures below approximately 1~K~\citep{dyakonov1972dynamic}. Below, we reconsider the Dyakonov and Perel model for the conditions of our experiments, and show that the DSPNS effect is a likely explanation of the observed features of the Hanle curves in Fig.~\ref{fig:Hanle_60_90_deg_Temperature}(c,f).

It has been shown previously that the spin polarization of lattice nuclei in GaAs follow the spin-temperature theory both at low~\citep{vladimirova2018spin} and high magnetic fields~\citep{chekhovich2017measurement}. Within the spin-temperature approach, the Overhauser field $\textbf{B}_{N}$ acting on electrons is parallel to the external magnetic field $\textbf{B}$ and its value is determined by the nuclear spin temperature $\Theta_{N}$
\begin{equation}
B_N=B_N^{max} {\rm B}_{3/2}\left(\frac{\hbar \bar{\gamma}_N B}{k_B \Theta_N}\right)\,.
\label{eq:nuclear_field}
\end{equation}
Here, $\rm B_{3/2}$ is the Brillouin function for spin $I=3/2$, $\bar{\gamma}_N$ is the nuclear gyromagnetic ratio averaged over isotopes, and $B_N^{max}$ is the Overhauser field at full polarization of nuclear spins. The spin temperature of nuclei coupled by the hyperfine interaction with photoexcited  electrons is given by ~\citep{dyakonov1975cooling}
\begin{equation}
\begin{split}
\frac{1}{k_B\Theta_N} &=\frac{4f}{\hbar\bar{\gamma}_N}\frac{B}{B^2+\tilde{B}_L^2}\frac{\bar{S}_B-S_T}{1-4\bar{S}_BS_T} \\
&\approx -\frac{4f}{\hbar\bar{\gamma}_N}\frac{B}{B^2+\tilde{B}_L^2}S_T.
\label{eq:spin_temperature}
\end{split}
\end{equation}
Here, $\tilde{B}_L$ is a characteristic field of nuclear spin-spin interactions, $\bar{S}_B$ is the projection of mean electron spin on the external field $\textbf{B}$ (on average, close to zero under modulated pumping) and  $f$ is a leakage factor. $S_T$ is the thermodynamically equilibrium value of the electron spin projection on $\textbf{B}$, determined by the total field acting upon electron spins and by the lattice temperature $T$
\begin{equation}
\begin{split}
S_T &= -\frac{1}{2} {\rm B}_{1/2}
    \left( \frac{\mu_B g_e (B_N + B)}{k_B T} \right) \\
    &= -\frac{1}{2} \tanh \left(
        \frac{\mu_B g_e (B_N + B)}{2 k_B T}
    \right)\,.
\label{eq:S_T_projection}
\end{split}
\end{equation}
Combining Eqs.~\eqref{eq:nuclear_field}, \eqref{eq:spin_temperature} and \eqref{eq:S_T_projection}, one arrives to an equation for the Overhauser field $B_N$ 
\begin{equation}
B_N = B_N^{max} {\rm B}_{3/2} \left[
    \frac{2 B^2 f}{B^2 + \tilde{B}_L^2}
    {\rm B}_{1/2}\left(
        \frac{\mu_B g_e (B_N + B)}{k_B T}
    \right)
\right].
\label{eq:nuclear_field_2}
\end{equation}
In order to analyze the behavior of the Overhauser field near zero external field, one can neglect $B$ as compared to $B_N$ in Eq.~\eqref{eq:nuclear_field_2}. Assuming that $\mu_Bg_eB_N \ll k_BT$, which is acceptable under our experimental conditions up to $B_N\sim 1$~T, we expand the Brillouin functions up to the 3-rd power of their arguments ${\rm B}_I(x) \approx \frac{1}{3}I(I+1)x - \frac{1}{90}I(I+1)[2I(I+1)-1]x^3$. These simplifications result in a 3-rd order algebraic equation for $z=\mu_Bg_eB_N/k_BT$
\begin{equation}
\begin{split}
\left(\frac{T}{T_0}\right)z &= \frac{5}{6} \frac{B^2f}{B^2 + B_L^2} \\
&\quad\times \left\{ z - \left[ \frac{1}{24} + \frac{17}{30} \left( \frac{B^2f}{B^2 + B_L^2} \right)^2 \right] z^3 \right\},
\label{eq:low-temperature_expansion}
\end{split}
\end{equation}
where $T_0=\mu_Bg_eB_N^{max}/k_B$. The value of $\Delta E_{eN}=\mu_Bg_eB_N^{max}$, which is the Overhauser splitting of electron spin levels under full polarization of nuclear spins, has been recently measured experimentally in GaAs~\citep{millington2024approaching} and equals approximately to 100~$\mu$eV, yielding $T_0 \approx 1.16$~K. 

Equation~\eqref{eq:low-temperature_expansion} has a trivial solution $z_0=0$, corresponding to the absence of nuclear polarization. However, if the temperature is lower than $T_c=\frac{5}{6}fT_0$, which is equal to 0.95~K in case of low leakage (where $T_{1e} \ll T_L$ and $f=1$), two other solutions appear in external magnetic fields exceeding the critical field $B_c=\tilde{B}_L\sqrt{\frac{T}{T_c-T}}$: 
\begin{equation}
z_{\pm} = \pm \sqrt{\frac{1 - \frac{6}{5} \frac{B^2 + \tilde{B}_L^2}{B^2 f} \left( \frac{T}{T_0} \right)}{\frac{1}{24} + \frac{17}{30} \left( \frac{B^2 f}{B^2 + \tilde{B}_L^2} \right)^2}}.
\label{eq:z_plus_minus}
\end{equation}
Besides, it can be shown that these states are stable, while $z_0$ loses stability. These solutions correspond to a non-zero Overhauser field, equal in absolute value to
\begin{equation}
\begin{split}
B_{Nsp}(B,T) &= B_N^{max} \left(\frac{T}{T_0}\right) |z_{\pm}| \\
&= B_N^{max} \left(\frac{T}{T_0}\right) \sqrt{\frac{1 - \frac{6}{5} \frac{B^2 + \tilde{B}_L^2}{B^2 f} \left( \frac{T}{T_0} \right)}{\frac{1}{24} + \frac{17}{30} \left( \frac{B^2 f}{B^2 + \tilde{B}_L^2} \right)^2}}.
\label{eq:B_N DSPNS}
\end{split}
\end{equation}
Appearance of these solutions manifests the onset of the DSPNS effect.

\begin{figure*}
\center{\includegraphics{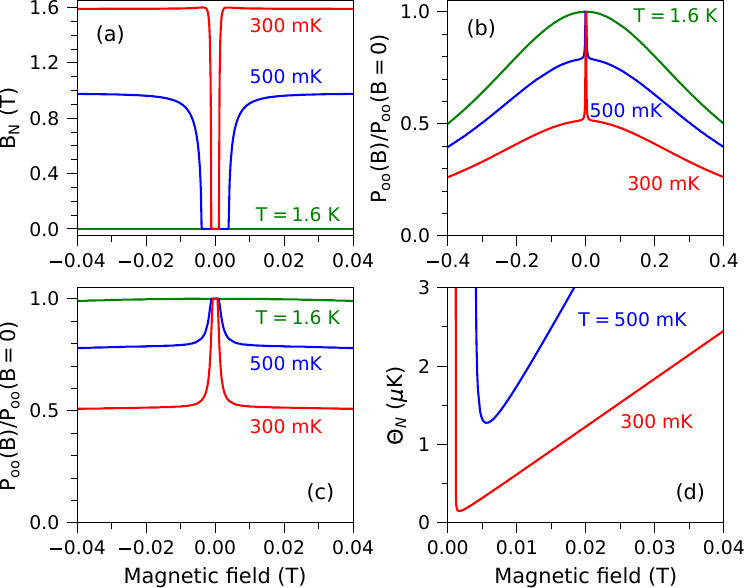}}
\caption{Model calculations of the Overhauser field $B_N$, arising as a result of DSPNS in the 19.7-nm-wide GaAs/Al$_{0.35}$Ga$_{0.65}$As QW, as a function of the external magnetic field $B$. $B_N$ is calculated using Eq.~\eqref{eq:B_N DSPNS} with the following parameters: $\tilde{B}_L=1$~mT and $B_{N}^{max}=4.3$~T. (a) Overhauser field as a function of the external magnetic field for $T=1.6$~K, 500~mK and 300~mK, and  for $f=0.55$. (b) Normalized PL polarization as a function of the external magnetic field for $T=1.6$~K, 500~mK and 300~mK ($f$ distributed from 0 to 1 with the maximum at 0.26). (c) Zoomed-in view of panel (b) in the magnetic-field range of $\pm0.04$~T. (d) Dependence of the nuclear spin temperature $\Theta_N$ on the external magnetic field for temperatures of 500~mK and 300~mK calculated for $f=0.55$.}
\label{fig:Theory_graphs}
\end{figure*}

One can see from Fig.~\ref{fig:Theory_graphs}(a) that, once self-polarization conditions are reached, the Overhauser field rapidly grows up to the values of about  1~T for $T=500$~mK and 1.6~T for $T=300$~mK. This results in the sharp drop in the PL polarization dependence on the transverse external magnetic field, see Fig.~\ref{fig:Theory_graphs}(b). That occurs when the external magnetic field $B$ reaches $B_c$ in absolute value, forming the zero-field peak in the Hanle curve, which is shown in more details in Fig.~\ref{fig:Theory_graphs}(c). Qualitatively, this behavior is very similar to what is observed experimentally, see Fig.~\ref{fig:Hanle_60_90_deg_Temperature}(f). 

However, the DSPNS theory faces some difficulties in quantitative description of our experimental results. First, the value of $f\approx0.26$, estimated from the oblique-field Hanle effect (Fig.~\ref{fig:hanle_tilted}) yields the value of $T_c$ slightly smaller than 300~mK,  so that the DSPNS is not expected to develop. Second, with the HWHM of the Hanle curve of 0.4~mT, the Overhauser field of the order of 1~T, which is predicted to develop once DNSPS occurs, would result in the polarization drop almost to zero, which is not observed in Fig.~\ref{fig:Hanle_60_90_deg_Temperature}(c,f). 

A possible explanation is the inhomogeneity of the leakage factor $f$, determined by the relation of relaxation times of nuclei via hyperfine interaction with electrons and other (presumably, quadrupole-induced~\citep{vladimirova2017nuclear}) relaxation mechanisms. Assuming that $f$ is statistically distributed in the range from 0 to 1 with the maximum probability near 0.26, and averaging the PL polarization in the transverse geometry over $f$ yields Hanle curves, similar to those observed experimentally. Compare, the Hanle curves in Fig.~\ref{fig:Theory_graphs}(b)  calculated using a truncated Gaussian distribution of $f$ with $\Delta f = 0.3$ and maximum at $f=0.26$ with experiments in Fig.~\ref{fig:Hanle_60_90_deg_Temperature}(e,f).

It is worth to summarize the required conditions for the DSPNS effect: 
\begin{itemize}
    \item Presence of charge carriers coupled by the hyperfine interaction with spins of lattice nuclei. 
   \item  Optical pumping with unpolarized/linearly polarized/polarization-modulated light or another external perturbation that would maintain the spins of charge carriers in the unpolarized state. 
   \item   Lattice temperature below some critical value (typically, in the subkelvin range), which depends on the leakage factor determined by the ratio of nuclear relaxation rates by hyperfine interaction with carriers and by other mechanisms. 
   \item   External magnetic field above certain critical value determined by local fields of nuclear spin-spin interactions and temperature (typically, of the order of 1~mT). The field direction can be arbitrary oriented, nuclear spins would polarize along this field.
  \end{itemize}
As all these conditions are satisfied in our experiments, we are firm that the observation of the sharp zero-field feature in the Hanle curve on lowering temperature below 1K can be interpreted as a manifestation of the DSPNS effect. 

\section{Conclusions}

Time-resolved Kerr rotation is used to study nuclear spin polarization and spin temperature in a 19.7-nm-wide GaAs quantum well. Long-term optical pumping with circularly polarized light produces an Overhauser field of up to 3.1~T at a $\theta=15^\circ$ tilt from the growth axis, corresponding to 72\% of the maximum achievable nuclear polarization in GaAs QWs. Given the relatively weak confinement of such a wide quantum well, this demonstrates a remarkably efficient nuclear spin polarization. Reaching this level constitutes a significant step towards achieving nuclear spin ordering, which is expected to emerge once the initial nuclear polarization approaches approximately 70\%.

Measurements of the nuclear spin temperature yield a minimum value of $\Theta_N=6.4$~$\mu$K at $B=0.006$~T, with extrapolation to zero field giving $\Theta_N=0.8$~$\mu$K. This value is slightly higher than the previously reported value of 0.54~$\mu$K on the same sample~\cite{kotur2021ultra}. 

At millikelvin lattice temperatures, nuclear spin dynamics is also examined. Despite the low optical power used to avoid heating, an Overhauser field of 0.16~T is achieved in an oblique (close to Voigt) field geometry. The nuclear build-up time of about 150~s is measured, being in good agreement with previously reported values at higher temperatures, where the hyperfine scattering on free photoexcited two-dimensional electrons is known to dominate the relaxation. Below approximately 6~K, however, the $T_1$ time becomes temperature dependent, most likely due to electron localization, which increases their correlation time with the nuclei and thereby enhances the hyperfine interaction.

On lowering the temperature to 500~mK and further to 300~mK, a sharp zero-field feature develops in the Hanle curve, which we interpret as evidence of the dynamic self-polarization of nuclear spins, predicted theoretically by Dyakonov and Perel in 1972~\citep{dyakonov1972dynamic}.

The presented results demonstrate that a wide undoped GaAs quantum well can sustain strong nuclear polarization and deep nuclear cooling, approaching the regime where collective nuclear-spin ordering is expected, paving the way for future studies of ordered nuclear-spin phases in semiconductors.

\section*{Acknowledgments} 
This study was made possible through funding of the He3-He4 dilution refrigerator through a Major Research Instrumentation Proposal of the Deutsche Forschungsgemeinschaft (project number 427377618).
K.V.K. acknowledges support by the Saint-Petersburg State University (grant No. 125022803069-4).


\begin{thebibliography}{99}

\bibitem{zutic2004spintronics}
I.~Žutić, J.~Fabian, and S.~Das Sarma,
Spintronics: Fundamentals and applications,
\href{https://doi.org/10.1103/RevModPhys.76.323}{Rev. Mod. Phys. \textbf{76}, 323 (2004)}.

\bibitem{kotur2021ultra}
M.~Kotur, D.~O.~Tolmachev, V.~M.~Litvyak, K.~V.~Kavokin, D.~Suter, D.~R.~Yakovlev, and M.~Bayer,
Ultradeep optical cooling of coupled nuclear spin-spin and quadrupole reservoirs in a GaAs/(Al,Ga)As quantum well,
\href{https://doi.org/10.1038/s42005-021-00681-6}{Commun. Phys. \textbf{4}, 193 (2021)}.

\bibitem{meier1984optical}
\textit{Optical Orientation}, edited by F.~Meier and B.~P.~Zakharchenya (North-Holland, Amsterdam, 1984).

\bibitem{witzel2007nuclear}
W.~M.~Witzel and S.~Das Sarma,
Nuclear spins as quantum memory in semiconductor nanostructures,
\href{https://doi.org/10.1103/PhysRevB.76.045218}{Phys. Rev. B \textbf{76}, 045218 (2007)}.

\bibitem{morton2008solid}
J.~J.~L.~Morton, A.~M.~Tyryshkin, R.~M.~Brown, S.~Shankar, B.~W.~Lovett, A.~Ardavan, T.~Schenkel, E.~E.~Haller, J.~W.~Ager, and S.~A.~Lyon,
Solid-state quantum memory using the $^{31}$P nuclear spin,
\href{https://doi.org/10.1038/nature07295}{Nature \textbf{455}, 1085 (2008)}.

\bibitem{abragam1961principles}
A.~Abragam, \textit{The Principles of Nuclear Magnetism} (Oxford University Press, Oxford, UK, 1961).

\bibitem{goldman1970spin}
M.~Goldman, \textit{Spin Temperature and Nuclear Magnetic Resonance in Solids} (Clarendon Press, UK, 1970)

\bibitem{merkulov1982phase}
I.~A.~Merkulov,
Phase transition in a nuclear spin system in a semiconductor with optically oriented electrons,
Zh. Eksp. Teor. Fiz. \textbf{82}, 319 (1982) [Sov. Phys. JETP \textbf{55}, 188 (1982)].

\bibitem{oja1997nuclear}
A.~S.~Oja, and O.~V.~Lounasmaa,
Nuclear magnetic ordering in simple metals at positive and negative nanokelvin temperatures,
\href{https://doi.org/10.1103/RevModPhys.69.1}{Rev. Mod. Phys. \textbf{69}, 1 (1997)}.

\bibitem{merkulov_polaron}
I.~A.~Merkulov,
Formation of a nuclear spin polaron under optical orientation in GaAs-type semiconductors,
Phys. Solid State \textbf{40}, 930 (1998).

\bibitem{vladimirova_polaron}
M.~Vladimirova, D.~Scalbert, M.~S.~Kuznetsova and K.~V.~Kavokin
Electron-induced nuclear magnetic ordering in n-type semiconductors,
Phys. Rev. B \textbf{103}, 205207 (2021).

\bibitem{mocek2017high}
R.~W.~Mocek, V.~L.~Korenev, M.~Bayer, M.~Kotur, R.~I.~Dzhioev, D.~O.~Tolmachev, G.~Cascio, K.~V.~Kavokin, and D.~Suter,
High-efficiency optical pumping of nuclear polarization in a GaAs quantum well,
\href{https://doi.org/10.1103/PhysRevB.96.201303}{Phys. Rev. B \textbf{96}, 201303 (2017)}.

\bibitem{kotur2018single}
M.~Kotur, F.~Saeed, R.~W.~Mocek, V.~L.~Korenev, I.~A.~Akimov, A.~S.~Bhatti, D.~R.~Yakovlev, D.~Suter, and M.~Bayer,
Single-beam resonant spin amplification of electrons interacting with nuclei in a GaAs/(Al,Ga)As quantum well,
\href{https://doi.org/10.1103/PhysRevB.98.205304}{Phys. Rev. B \textbf{98}, 205304 (2018)}.


\bibitem{eickhoff2002coupling}
M.~Eickhoff, B.~Lenzmann, G.~Flinn, and D.~Suter,
Coupling mechanisms for optically induced NMR in GaAs quantum wells,
\href{https://doi.org/10.1103/PhysRevB.65.125301}{Phys. Rev. B \textbf{65}, 125301 (2002)}.

\bibitem{lampel1968nuclear}
G.~Lampel,
Nuclear dynamic polarization by optical electronic saturation and optical pumping in semiconductors,
\href{https://doi.org/10.1103/PhysRevLett.20.491}{Phys. Rev. Lett. \textbf{20}, 491 (1968)}.

\bibitem{dyakonov2017spin}
\textit{Spin Physics in Semiconductors}, edited by M.~I.~Dyakonov (Springer International Publishing AG, 2017).

\bibitem{worsley1996transient}
R.~E.~Worsley, N.~J.~Traynor, T.~Grevatt, and R.~T.~Harley,
Transient linear birefringence in GaAs quantum wells: Magnetic field dependence of coherent exciton spin dynamics,
\href{https://doi.org/10.1103/PhysRevLett.76.3224}{Phys. Rev. Lett. \textbf{76}, 3224 (1996)}.

\bibitem{kikkawa2000all}
J.~M.~Kikkawa and D.~D.~Awschalom,
All-optical magnetic resonance in semiconductors,
\href{https://doi.org/10.1126/science.287.5452.473}{Science \textbf{287}, 473 (2000)}.

\bibitem{snelling1991magnetic}
M.~J.~Snelling, G.~P.~Flinn, A.~S.~Plaut, R.~T.~Harley, A.~C.~Tropper, R.~Eccleston, and C.~C.~Phillips,
Magnetic g factor of electrons in GaAs/Al$_x$Ga$_{1-x}$As quantum wells,
\href{https://doi.org/10.1103/PhysRevB.44.11345}{Phys. Rev. B \textbf{44}, 11345 (1991)}.

\bibitem{aleksandrov2011spin}
E.~B.~Aleksandrov and V.~S.~Zapasskii,
Spin noise spectroscopy,
\href{https://doi.org/10.1088/1742-6596/324/1/012002}{J. Phys.: Conf. Ser. \textbf{324}, 012002 (2011)}.

\bibitem{baugh2007large}
J.~Baugh, Y.~Kitamura, K.~Ono, and S.~Tarucha,
Large nuclear Overhauser fields detected in vertically coupled double quantum dots,
\href{https://doi.org/10.1103/PhysRevLett.99.096804}{Phys. Rev. Lett. \textbf{99}, 096804 (2007)}.

\bibitem{tartakovskii2007nuclear}
A.~I.~Tartakovskii, T.~Wright, A.~Russell, V.~I.~Fal'ko, A.~B.~Van'kov, J.~Skiba-Szymanska, I.~Drouzas, R.~S.~Kolodka, M.~S.~Skolnick, P.~W.~Fry, A.~Tahraoui, H.-Y.~Liu, and M.~Hopkinson,
Nuclear spin switch in semiconductor quantum dots,
\href{https://doi.org/10.1103/PhysRevLett.98.026806}{Phys. Rev. Lett. \textbf{98}, 026806 (2007)}.

\bibitem{millington2024approaching}
P.~Millington-Hotze, H.~E.~Dyte, S.~Manna, S.~F.~Covre~da~Silva, A.~Rastelli, and E.~A.~Chekhovich,
Approaching a fully-polarized state of nuclear spins in a solid,
\href{https://doi.org/10.1038/s41467-024-45364-2}{Nat. Commun. \textbf{15}, 985 (2024)}.

\bibitem{vladimirova2018spin}
M.~Vladimirova, S.~Cronenberger, D.~Scalbert, I.~I.~Ryzhov, V.~S.~Zapasskii, G.~G.~Kozlov, A.~Lemaître, and K.~V.~Kavokin,
Spin temperature concept verified by optical magnetometry of nuclear spins,
\href{https://doi.org/10.1103/PhysRevB.97.041301}{Phys. Rev. B \textbf{97}, 041301 (2018)}.

\bibitem{paget1977low}
D.~Paget, G.~Lampel, B.~Sapoval, and V.~I.~Safarov,
Low-field electron-nuclear spin coupling in gallium arsenide under optical pumping conditions,
\href{https://doi.org/10.1103/PhysRevB.15.5780}{Phys. Rev. B \textbf{15}, 5780 (1977)}.

\bibitem{vladimirova2017nuclear}
M.~Vladimirova, S.~Cronenberger, D.~Scalbert, M.~Kotur, R.~I.~Dzhioev, I.~I.~Ryzhov, G.~G.~Kozlov, V.~S.~Zapasskii, A.~Lemaître, and K.~V.~Kavokin,
Nuclear spin relaxation in n-GaAs: From insulating to metallic regime,
\href{https://doi.org/10.1103/PhysRevB.95.125312}{Phys. Rev. B \textbf{95}, 125312 (2017)}.

\bibitem{gribakin2024nuclear}
B.~F.~Gribakin, V.~M.~Litvyak, M.~Kotur, R.~André, M.~Vladimirova, D.~R.~Yakovlev, and K.~V.~Kavokin,
Nuclear spin relaxation mediated by donor-bound and free electrons in wide CdTe quantum wells,
\href{https://doi.org/10.1103/PhysRevB.109.195302}{Phys. Rev. B \textbf{109}, 195302 (2024)}.

\bibitem{hanle1924magnetische}
W.~Hanle,
Über magnetische Beeinflussung der Polarisation der Resonanzfluoreszenz,
Z. Phys. \textbf{30}, 93 (1924).

\bibitem{d1974optical}
M.~I.~D'yakonov, V.~I.~Perel', V.~L.~Berkovits, and V.~I.~Safarov,
Optical effects due to polarization of nuclei in semiconductors,
Zh. Eksp. Teor. Fiz. \textbf{67}, 1912 (1974) [Sov. Phys. JETP \textbf{40}, 950 (1975)].


\bibitem{kalevich1982onset}
V.~K.~Kalevich, V.~D.~Kulkov, and V.~G.~Fleisher,
Onset of a nuclear polarization front due to optical spin orientation in a semiconductor,
Pis’ma Zh. Eksp. Teor. Fiz. \textbf{35}, 17 (1982) [JETP Lett. \textbf{35}, 20 (1982)].

\bibitem{kotur2014nuclear}
M.~Kotur, R.~I.~Dzhioev, K.~V.~Kavokin, V.~L.~Korenev, B.~R.~Namozov, P.~E.~Pak, and Y.~G.~Kusrayev,
Nuclear spin relaxation mediated by Fermi-edge electrons in n-type GaAs,
\href{https://doi.org/10.1134/S0021364014010068}{JETP Lett. \textbf{99}, 37 (2014)}.

\bibitem{kalevich1990optical}
V.~K.~Kalevich, V.~L.~Korenev, and O.~M.~Fedorova,
Optical polarization of nuclei in GaAs/AlGaAs quantum-well structures,
Pis’ma Zh. Eksp. Teor. Fiz. \textbf{52}, 964 (1990) [JETP Lett. \textbf{52}, 349 (1990)].

\bibitem{dyakonov1972dynamic}
M.~I.~D'yakonov, V.~I.~Perel',
Dynamic self-polarization of nuclei in solids,
Pis’ma Zh. Eksp. Teor. Fiz. \textbf{16}, 563 (1972) [JETP Lett. \textbf{16}, 398 (1972)].

\bibitem{chekhovich2017measurement}
E.~A.~Chekhovich, A.~Ulhaq, E.~Zallo, F.~Ding, O.~G.~Schmidt, and M.~S.~Skolnick,
Measurement of the spin temperature of optically cooled nuclei and GaAs hyperfine constants in GaAs/AlGaAs quantum dots,
\href{https://doi.org/10.1038/nmat4959}{Nat. Mater. \textbf{16}, 982 (2017)}.

\bibitem{dyakonov1975cooling}
M.~I.~D'yakonov and V.~I.~Perel',
Cooling of a system of nuclear spins following optical orientation of electrons in semiconductors,
Zh. Eksp. Teor. Fiz. \textbf{68}, 1514 (1975) [JETP \textbf{41}, 759 (1975)].

\end{thebibliography}
\end{document}